\documentclass[conference]{IEEEtran}
\IEEEoverridecommandlockouts

\usepackage{cite}
\usepackage{marvosym}
\usepackage{amsmath,amssymb,amsfonts}
\usepackage{graphicx}
\usepackage{booktabs}
\usepackage{multirow}
\usepackage{array}
\usepackage{makecell}
\usepackage[table]{xcolor}
\definecolor{conciserow}{gray}{0.92}
\newcommand{\up}[1]{\textcolor{green!60!black}{\textbf{+#1}}}
\newcommand{\down}[1]{\textcolor{red!70!black}{--#1}}
\usepackage{algorithm}
\makeatletter
\let\COMMENT\@undefined
\makeatother
\usepackage{algorithmic}
\usepackage{textcomp}
\usepackage{pifont}
\usepackage{hyperref}
\hypersetup{
    colorlinks=true,
    linkcolor=blue,
    filecolor=blue,      
    urlcolor=blue,
    citecolor=blue,
}

\begin{document}

\title{ConCise: Training-Free Conclusion-Chain State Compression for Cost-Efficient \\Multi-Step RAG Services}

\author{
\thanks{This work was supported in part by the National Natural Science Foundation of China (NSFC) under Grant 62302048, Grant U25A20436, and Grant 62272050; in part by Guangdong Higher Education Association under Grant 24GQN97; in part by the Guangdong Provincial Higher Education Institutions under Grant 2024KTSCX219; and in part by Beijing Normal University at Zhuhai Education Reform Project under Grant jx2025037. \textit{(Corresponding author: Zhiqing Tang.)}}
\IEEEauthorblockN{
Kuan Yan\textsuperscript{1,2},
Zhiqing Tang\textsuperscript{2,\Letter},
Tian Wang\textsuperscript{2}, and
Weijia Jia\textsuperscript{2,3}
}

\IEEEauthorblockA{
\textsuperscript{1}Faculty of Arts and Sciences, Beijing Normal University, Zhuhai, China\\
\textsuperscript{2}Institute of Artificial Intelligence and Future Networks, Beijing Normal University, Zhuhai, China\\
\textsuperscript{3}Guangdong Key Lab of AI \& Multi-Modal Data Processing, Beijing Normal-Hong Kong Baptist University, Zhuhai, China\\
yankuan@mail.bnu.edu.cn, \{zhiqingtang, tianwang, jiawj\}@bnu.edu.cn
}
}

\maketitle
\begin{abstract}
Multi-step retrieval-augmented generation (RAG) has been widely deployed as LLM-powered web services for complex question answering, where iterative retrieval-reasoning rounds deliver strong multi-hop accuracy. However, this paradigm causes historical documents and reasoning traces to accumulate across rounds, inflating cumulative input tokens approximately as $O(N^2)$ with progressively increasing noise density. In API-based service architectures, such growth directly amplifies per-request billing cost, network payload, and response latency. Existing compression approaches rely on pretrained modules or GPU-level KV cache access, introducing model hosting overhead incompatible with API-native, Serverless, and edge-side deployments. To address this issue, this paper proposes ConCise, a training-free state-layer protocol that restructures cross-round context transmission for multi-step RAG services. Specifically, ConCise replaces raw-text accumulation with an append-only chain of structured conclusions, compressing cumulative context growth from $O(N^2)$ to approximately $O(N)$. Furthermore, a fused generation mechanism is introduced to jointly emit reasoning and conclusions in a single API call, eliminating repeated input billing from serial dual-invocation overhead. Extensive experiments across twelve paired configurations spanning three models, two datasets, and two representative frameworks demonstrate that ConCise achieves 64.63\% average token savings while maintaining acceptable accuracy, providing a plug-and-play, deployment-friendly solution for cost-efficient multi-step RAG service optimization.

\end{abstract}


\section{Introduction}

Retrieval-augmented generation (RAG)~\cite{lewis2020rag} has become a core technique in large language model (LLM) applications, widely used to reduce hallucinations and improve factual accuracy. With the growth of LLM-as-a-Service, RAG is increasingly delivered through RESTful APIs as part of cloud-native Web services for tasks such as enterprise knowledge retrieval and intelligent assistants. In these API-based settings, each request is billed by input token count and also incurs network and latency overhead, so token efficiency is a key service-level concern. As tasks grow more complex, single-step retrieval is often not enough for multi-hop questions. This has led RAG systems to adopt multi-step designs that alternate between retrieval and reasoning over several rounds. Frameworks such as IRCoT~\cite{trivedi2023ircot}, FLARE~\cite{jiang2023flare}, Self-RAG~\cite{asai2024selfrag}, and Search-R1~\cite{jin2025searchr1} show strong results on complex question answering with this iterative approach.

\begin{figure}[t]
\centering
  \includegraphics[width=\columnwidth]{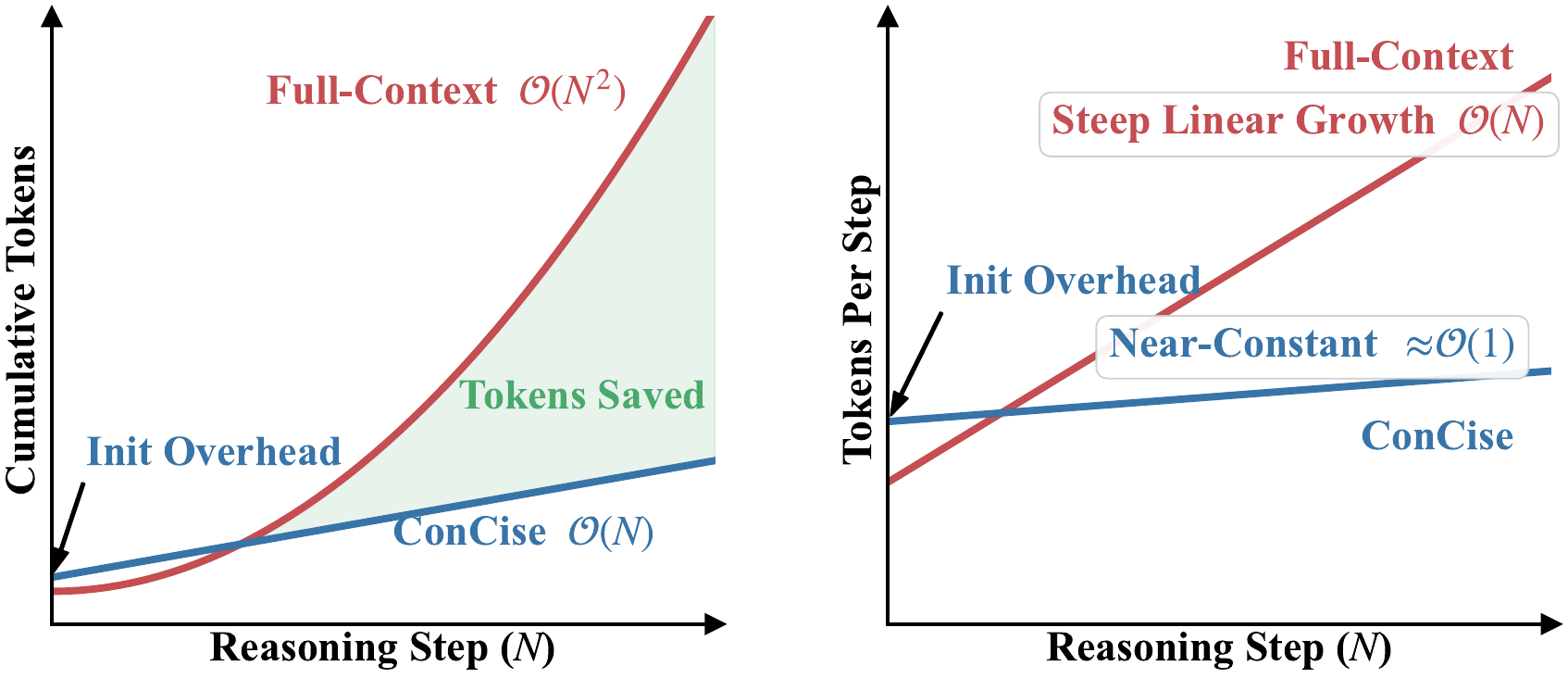}
\caption{Token consumption trend comparison. In the multi-step retrieval-inference loop, the traditional full-context paradigm shows cumulative growth. ConCise, by only propagating the conclusion chain and discarding historical raw text, compresses the cumulative growth from approximately $O(N^2)$ to approximately $O(N)$.}
\label{fig:token_trend}
\end{figure}

However, it also brings historical context inflation. Under the full-context paradigm, all retrieved documents and reasoning traces are carried into every subsequent round, causing cumulative input tokens to grow as roughly $O(N^2)$ while noise density keeps increasing~\cite{liu2024lostmiddle}(Fig.~\ref{fig:token_trend}). In API-based services, this growth leads directly to higher per-request cost, larger payloads, and longer response times. Existing compression methods treat multi-step RAG as a static long-document problem. They are not built for the round-by-round, append-only nature of multi-hop reasoning. Despite the wide use of multi-step RAG in web services and API-based deployments, no training-free solution targets this dynamic setting. Current methods mostly require offline training or model access, leaving API-native, serverless, and edge deployments without a lightweight option for iterative state management. 
Recent work has explored infrastructure-level QoS optimization for LLM or AIGC services~\cite{cloudedge2024llmqos,eat2024qos,velo2024llm,mou2026adaptive,li2025cloudedge,sheng2026adapting}, however, these works do not address prompt-level state compression for iterative RAG.
Therefore, a training-free approach that compresses multi-step RAG state at low cost while preserving reasoning quality is needed. To this end, two challenges must be addressed.

\textit{The first challenge is to compress cross-round state with low token cost while retaining the multi-hop evidence needed for accurate answers.} Existing methods fall into four categories: (1)~token-level pruning (LLMLingua~\cite{jiang2023llmlingua}, EXIT~\cite{hwang2025exit}); (2)~KV cache optimization (SnapKV~\cite{li2024snapkv}); (3)~soft vector compression (ICAE, AutoCompressor, xRAG~\cite{ge2023icae,chevalier2023autocompressor,cheng2024xrag}); and (4)~recursive summarization (RECOMP~\cite{xu2023recomp}). None of them is built for the iterative nature of multi-step RAG. Token-level and KV-cache methods lack semantic modeling and may drop weakly linked early clues that later reasoning depends on. Soft compression needs fine-tuning. Recursive summarization recompresses earlier states at every round, so key entities, constraints, and relations found in early rounds can be gradually lost, degrading answer accuracy rather than only cross-round coherence. To address this, we replace raw-text accumulation with an append-only Chain of Conclusions. Each round keeps only a compact conclusion sequence, reducing cumulative growth from $O(N^2)$ to $O(N)$ while retaining the decision-relevant facts needed for multi-hop reasoning.

\textit{The second challenge is to implement such compression without model access or additional training, keeping it compatible with black-box API deployments.} KV cache methods need GPU-level model access, which is not available through black-box APIs. Soft compression requires a separate model, creating a two-model pipeline with extra orchestration and cold-start cost. Training-based methods need fine-tuning, which is impractical for thin-client or edge nodes. In addition, compression strategies that require extra serial API calls may offset part of the token savings through repeated input billing and latency. 
Even recent agentic frameworks such as ReAct~\cite{yao2023react} and Search-R1, as well as prompt-driven methods like IRCoT, still carry full raw text across rounds.
Currently, no training-free method compresses iterative RAG state while remaining compatible with black-box APIs. To fill this gap, we design ConCise as a training-free state management method that works at the prompt-concatenation level. It also introduces fused generation, which produces reasoning and a structured conclusion in one pass, removing the repeated input cost of the standard two-call workflow.

In this paper, we propose ConCise, a training-free conclusion-chain state management method for cost-efficient multi-step RAG services. We evaluate ConCise through 12~sets of paired experiments on 2WikiMultihopQA~\cite{ho2020multihop} and HotpotQA~\cite{yang2018hotpotqa}, using three models (GPT-5, Qwen2.5-72B, DeepSeek~V3.2) with IRCoT~\cite{trivedi2023ircot} and Search-R1~\cite{jin2025searchr1} as baselines. The maximum reasoning depth is set to~4, matching the budget of mobile and edge service deployments. The main contributions are summarized as follows.

\begin{enumerate}
    \item \textbf{Conclusion-chain iterative paradigm.} We replace raw-text accumulation with an append-only chain of structured conclusions, reducing $O(N^2)$ context growth to $O(N)$ and removing cross-round noise propagation. We provide an information-theoretic analysis showing that this append-only rule avoids the cascading information loss of recursive summarization.
    \item \textbf{Training-free pluggable state architecture with fused generation.} We implement this as a zero-training method that needs no model modification, and add a fused generation mode that produces reasoning and conclusions in a single API call, cutting per-step overhead for API-native and edge Web service deployments.
    \item \textbf{Systematic empirical evaluation.} Across all 12~configurations, ConCise achieves 64.63\% average token savings at step~4 while keeping accuracy loss within acceptable bounds, confirming the effectiveness of conclusion-chain state compression across diverse models and frameworks.
\end{enumerate}

The remainder of this paper is organized as follows. Section~\ref{sec:related} reviews related work. Section~\ref{sec:model} introduces the system model and problem formulation. Section~\ref{sec:algorithm} presents the ConCise algorithm. Section~\ref{sec:theory} provides theoretical analysis. Section~\ref{sec:experiment} reports experimental results and case studies. Finally, Section~\ref{sec:conclusion} concludes the paper.

\section{Related Work}
\label{sec:related}

\subsection{Iterative Reasoning and Multi-Step RAG}

Early work on deep reasoning focuses on prompt engineering, such as Chain-of-Thought (CoT)~\cite{wei2022cot}, Zero-shot CoT~\cite{kojima2022zeroshot}, and Tree-of-Thought (ToT)~\cite{yao2023tot}. RAG~\cite{lewis2020rag} extends these ideas from single-round retrieval to multi-step architectures that alternate between retrieval and reasoning. Representative frameworks include Self-Ask~\cite{khattab2022selfask}, ReAct~\cite{yao2023react}, IRCoT~\cite{trivedi2023ircot}, FLARE~\cite{jiang2023flare}, Self-RAG~\cite{asai2024selfrag}, Search-o1~\cite{li2025searcho1}, Search-R1~\cite{jin2025searchr1}, and DeepNote~\cite{wang2025deepnote}. These systems explore different strategies including problem decomposition, tool calling, and active retrieval. RAT~\cite{wang2024rat} improves single-round reasoning quality by incorporating retrieval results. MCoT and InftyThink~\cite{yang2025mcot,inftythink2025} show that iterative states can replace full text accumulation in long-context reasoning. However, all these frameworks carry raw documents and reasoning traces across rounds without compression. As the number of rounds grows, this leads to rapidly increasing token cost, which is especially problematic for API-billed web service deployments.

\subsection{Long Context Compression}

Existing compression methods can be grouped into four categories: (1)~token-level pruning (LLMLingua~\cite{jiang2023llmlingua}, EXIT~\cite{hwang2025exit}), which removes less important tokens from the input; (2)~system-level KV cache optimization (SnapKV~\cite{li2024snapkv}), which reduces memory usage at the attention layer; (3)~soft vector compression (ICAE~\cite{ge2023icae}, AutoCompressor~\cite{chevalier2023autocompressor}, xRAG~\cite{cheng2024xrag}), which encodes long contexts into compact learned representations; and (4)~recursive summarization (RECOMP~\cite{xu2023recomp}), which condenses documents through repeated summary passes. While effective for static long documents, these methods are not designed for multi-step RAG. Moreover, all of them require either model access or extra model hosting, making them incompatible with the black-box API settings common in Web service deployments. None offers a training-free solution for iterative state compression in multi-step RAG.

Beyond these categories, recent work applies token-level or attention-based pruning within RAG pipelines to reduce per-round input length. These methods and ConCise operate at different layers. Pruning methods compress a single retrieval context within one round, while ConCise compresses the historical state carried across rounds. The two are complementary: pruning can reduce the current document $\mathcal{D}_t$, while ConCise replaces accumulated history with a compact conclusion chain.

\section{System Model and Problem Formulation}
\label{sec:model}

\subsection{System Model}

We consider a multi-step RAG system that answers a complex question $q$ through $N$ retrieval-reasoning rounds. We use $\oplus$ to denote concatenation throughout this paper. In each round $t \in \{1,\dots,N\}$, the retriever $\mathcal{R}$ returns a document set $\mathcal{D}_t$ based on the current query and history. The generator then produces a reasoning trajectory $r_t$ and extracts a structured conclusion $c_t$.

The conclusion chain up to round $t$ is $\mathcal{C}_t = [c_1, c_2, \dots, c_t]$, growing by one entry per round. Under the full-context setting, the complete history is $\mathcal{H}_t = [\mathcal{D}_1, r_1, \dots, \mathcal{D}_t, r_t]$. Under ConCise, the compressed state $\mathcal{S}_t = (q, \mathcal{C}_t)$ replaces $\mathcal{H}_t$ as the cross-round carry-over. Let $\tau(\cdot)$ denote the token counting function. At each round $t$, the full-context setting keeps both $\mathcal{D}_t$ and $r_t$ for all future rounds, while ConCise keeps only $c_t$. The per-round token saving is:
\begin{equation}
\label{eq:delta}
\Delta_t = \tau(\mathcal{D}_t) + \tau(r_t) - \tau(c_t).
\end{equation}

\subsection{Problem Formulation}
 
Under the full-context setting, the input to round $t$ accumulates all previous documents and reasoning traces:
\begin{equation}
x^{\text{full}}_t = q \oplus [\mathcal{D}_1,r_1,\dots,\mathcal{D}_{t-1},r_{t-1},\mathcal{D}_t].
\end{equation}
This preserves complete information but causes cumulative input tokens to grow as roughly $O(N^2)$, with increasing noise density that leads to attention drift and degraded accuracy~\cite{liu2024lostmiddle}.
 
\textbf{Problem statement.} Let $\mathrm{Acc}(\cdot)$ denote the end-task answer accuracy. Given a multi-step RAG system with full-context inputs $\{x^{\text{full}}_t\}_{t=1}^{N}$, the goal is to find a compressed alternative $\{x^{\text{opt}}_t\}_{t=1}^{N}$ that keeps accuracy loss within a bound $\epsilon$ while reducing total token consumption:
\begin{equation}
\text{find } \{x^{\text{opt}}_t\}, \quad
\text{s.t.}
\begin{cases}
\mathrm{Acc}(\{x^{\text{full}}_t\}) - \mathrm{Acc}(\{x^{\text{opt}}_t\}) \leq \epsilon, \\[4pt]
\displaystyle\sum_{t=1}^{N} \tau(x^{\text{opt}}_t) \ll \sum_{t=1}^{N} \tau(x^{\text{full}}_t),
\end{cases}
\end{equation}
where accuracy is the primary constraint and token reduction is the optimization objective.

In API-based web service deployments, token consumption also determines two service-side costs. The estimated network payload at step $t$ is:
\begin{equation}
P_t = \frac{\tau(x^{\text{opt}}_t) \cdot b}{1024}, \quad b = 4\ \text{bytes/token},
\end{equation}
and the estimated input billing cost (in cents) is:
\begin{equation}
C_t = \tau(x^{\text{opt}}_t) \cdot \frac{p}{10^6} \cdot 100,
\end{equation}
where $p$ is the input token unit price (USD per 1M tokens). Both $P_t$ and $C_t$ are linear in $\tau(x^{\text{opt}}_t)$, so reducing token count proportionally reduces payload and cost. ConCise instantiates $x^{\text{opt}}_t$ as:
\begin{equation}
\label{eq:concise_input}
x^{\text{concise}}_t = q \oplus \mathcal{C}_{t-1} \oplus \mathcal{D}_t,
\end{equation}
where only the conclusion chain $\mathcal{C}_{t-1}$ crosses round boundaries, and $\mathcal{D}_t$ is used only within the current round. This reduces cumulative growth from roughly $O(N^2)$ to $O(N)$, with payload and billing cost scaling proportionally.
 
\begin{figure}[t]
\centering
\includegraphics[width=\columnwidth]{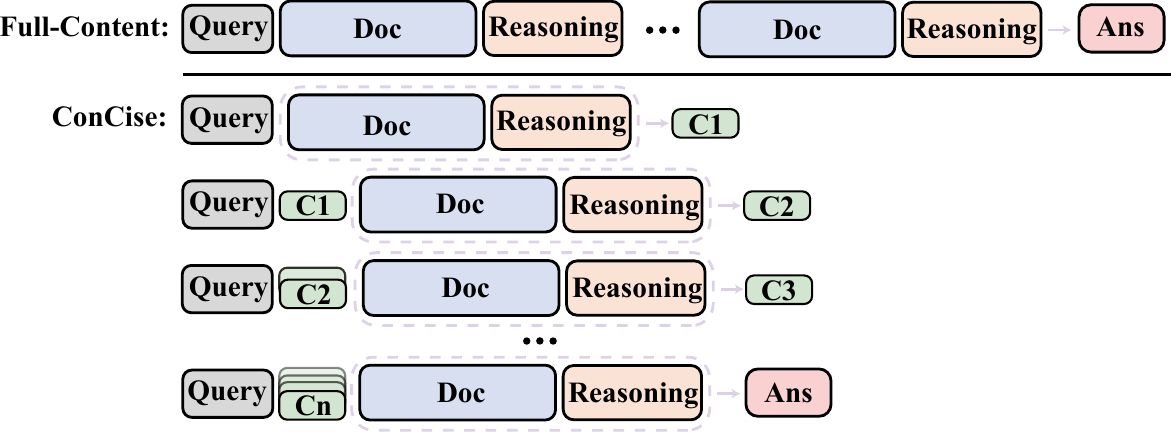}
\caption{Comparison of state evolution between full-context and ConCise. In full-context, each round accumulates documents and reasoning traces. In ConCise, each round appends only the current conclusion $c_t$ to the chain $\mathcal{C}_{t-1}$, which together with the current document forms the input for the next round.}
\label{fig:state_evolution}
\end{figure}
 
\begin{figure*}[t]
\centering
\includegraphics[width=0.83\textwidth]{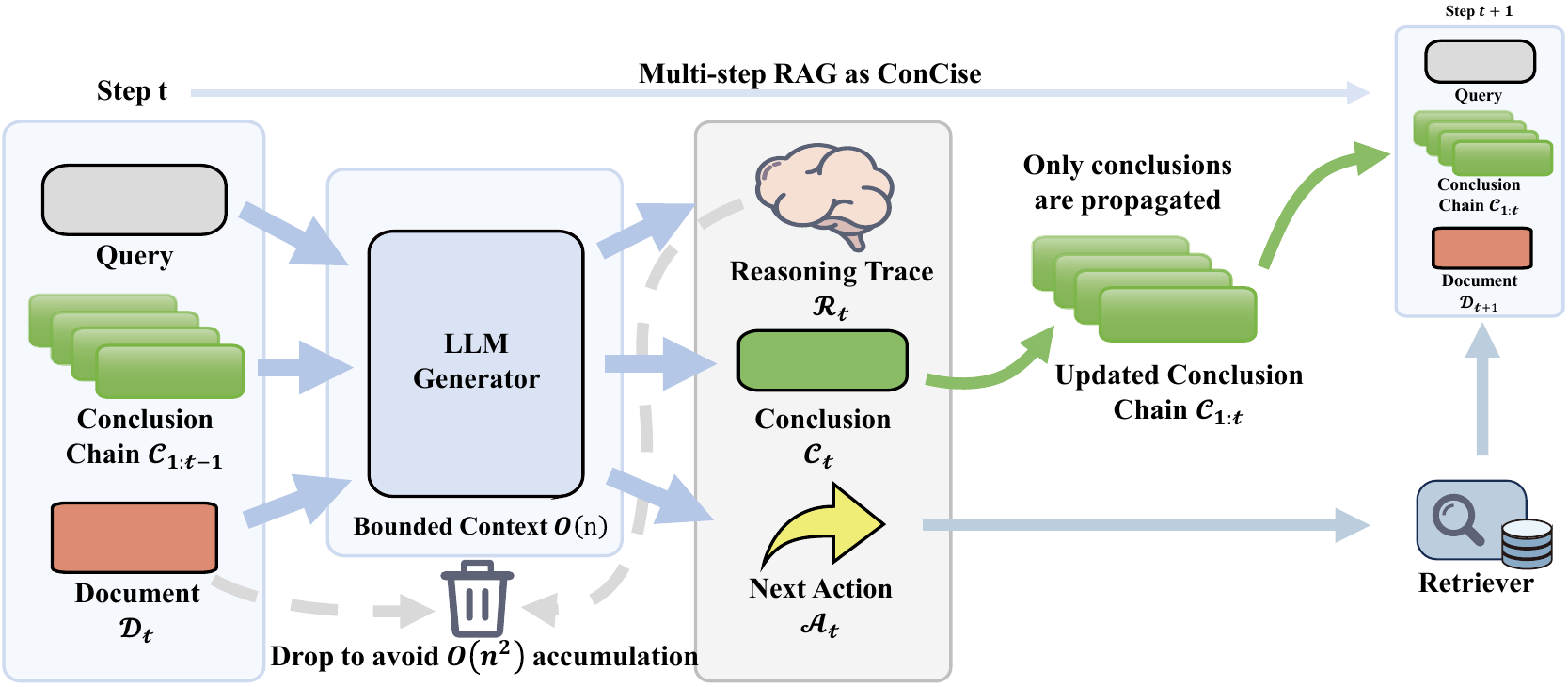}
\caption{Overview of the ConCise architecture. At step $t$, the system concatenates the query, the conclusion chain, and the current document into a bounded context. Through fused generation or a two-stage process of reasoning and summarization, it produces the reasoning trajectory, a structured conclusion, and the action for the next step. Only the conclusion chain propagates across steps; raw documents do not enter subsequent history.}
\label{fig:concise_arch}
\end{figure*}

\section{Algorithm}
\label{sec:algorithm}

\subsection{Conclusion-Chain State Iteration}

Algorithm~\ref{alg:concise_main} shows the main process of ConCise. Given a question $q$, the system runs up to $N$ retrieval-reasoning rounds. Each round has four steps. First, the retriever $\mathcal{R}$ fetches a document set $\mathcal{D}_t$ using the query and the current conclusion chain (line~3). Second, the system builds a bounded input $x_t$ by concatenating $q$, the conclusion chain $\mathcal{C}_{t-1}$, and $\mathcal{D}_t$ (line~4). No historical raw text is included. Third, a state update module $\mathcal{U}$ takes $x_t$ and produces a reasoning trajectory $r_t$, a structured conclusion $c_t$, and an action $a_t$ (line~5). Fourth, $c_t$ is appended to the chain (line~6). If $a_t$ indicates a final answer, the loop terminates early. The output is the final answer $a^*$ and the accumulated conclusion chain.

\begin{algorithm}[htbp]
\caption{ConCise}
\label{alg:concise_main}
{\raggedright
\textbf{Input:} Query $q$, max steps $N$, retriever $\mathcal{R}$, state update module $\mathcal{U}$ \\
\textbf{Output:} Final answer $a^*$, conclusion chain $\mathcal{C}$ \par}
\begin{algorithmic}[1]
\STATE Initialize conclusion chain $\mathcal{C}_0 \leftarrow \emptyset$
\FOR{$t = 1$ to $N$}
    \STATE $\mathcal{D}_t \leftarrow \mathcal{R}(q, \mathcal{C}_{t-1})$
    \STATE $x_t \leftarrow q \oplus \mathcal{C}_{t-1} \oplus \mathcal{D}_t$
    \STATE $(r_t, c_t, a_t) \leftarrow \mathcal{U}(x_t)$ \hfill \COMMENT{\textit{$\triangleright$ Call Algorithm~\ref{alg:concise_update}}}
    \STATE $\mathcal{C}_t \leftarrow \operatorname{Append}(\mathcal{C}_{t-1}, c_t)$
    \IF{$a_t$ is \textbf{ANSWER}($a^*$)} \STATE \textbf{break} \ENDIF
\ENDFOR
\RETURN $a^*, \mathcal{C}_t$
\end{algorithmic}
\end{algorithm}

We now explain three key design decisions. \textbf{Bounded state input (line~4).} Instead of carrying all previous documents and traces as in $x^{\text{full}}_t$, ConCise builds the input from only $q$, $\mathcal{C}_{t-1}$, and $\mathcal{D}_t$ (Eq.~(\ref{eq:concise_input})). The document $\mathcal{D}_t$ is used only in the current round; only $\mathcal{C}_{t-1}$ crosses round boundaries. Because each $c_t$ is much shorter than the pair $(\mathcal{D}_t, r_t)$ it replaces, the per-round saving $\Delta_t$ (Eq.~(\ref{eq:delta})) accumulates over rounds. Fig.~\ref{fig:state_evolution} and Fig.~\ref{fig:concise_arch} illustrate the state evolution and overall architecture.
 
\textbf{Joint conditional conclusion extraction (line~5).}
A simpler design would generate the conclusion from $q$ and $\mathcal{D}_t$ alone, but this ignores the reasoning trajectory $r_t$ and the prior chain $\mathcal{C}_{t-1}$, making it easy to lose intermediate-entity clues in multi-hop tasks. ConCise instead conditions on all four inputs:
\begin{equation}
c_t = \mathcal{U}_{\text{comp}}(\mathcal{D}_t, r_t \mid q, \mathcal{C}_{t-1}).
\end{equation}
The bar separates the factual source from the conditioning context. The query $q$ and prior chain $\mathcal{C}_{t-1}$ guide what to extract but are not written into $c_t$ as new facts.
 
\textbf{Append-only state transition (line~6).}
An alternative is to merge $c_t$ with previous conclusions into a single summary. However, repeated rewriting causes early facts to degrade, since each pass recompresses previously confirmed information. ConCise instead uses an append-only rule:
\begin{equation}
\mathcal{C}_t = \operatorname{Append}(\mathcal{C}_{t-1}, c_t), \quad \mathcal{S}_t = (q, \mathcal{C}_t).
\end{equation}
Each earlier conclusion $c_k$ ($k \le t$) is kept verbatim and never recompressed. Compared with the full history $\mathcal{H}_t$, ConCise carries only $\mathcal{S}_t$, with $|\mathcal{S}_t| \ll |\mathcal{H}_t|$ in practice, reducing context growth from roughly $O(N^2)$ to $O(N)$.

\subsection{Pluggable State Update Mechanism}

The state update module $\mathcal{U}$ in Algorithm~\ref{alg:concise_main} (line~5) takes the bounded input $x_t$ and produces a reasoning trajectory $r_t$, a structured conclusion $c_t$, and an action $a_t$. A straightforward implementation would use two sequential API calls: one for reasoning and one for conclusion extraction. However, the second call re-sends $x_t$ as input, doubling the per-step billing. To keep ConCise lightweight, we design $\mathcal{U}$ as a pluggable module with two interchangeable modes (Algorithm~\ref{alg:concise_update}).

\begin{algorithm}[htbp]
\caption{Pluggable State Update Module $\mathcal{U}$}
\label{alg:concise_update}
{\raggedright
\textbf{Input:} Bounded context $x_t$, mode, main model $\mathcal{M}_{\text{main}}$, summary model $\mathcal{M}_{\text{sum}}$ \\
\textbf{Output:} Reasoning trace $r_t$, structured conclusion $c_t$, action $a_t$ \par}
\begin{algorithmic}[1]
\IF{mode $=$ fused}
    \STATE $(r_t, c_t, a_t) \leftarrow \mathcal{M}_{\text{main}}(x_t)$
\ELSE
    \STATE $(r_t, a_t) \leftarrow \mathcal{M}_{\text{main}}(x_t)$ 
    \STATE $c_t \leftarrow \mathcal{M}_{\text{sum}}(x_t \oplus r_t)$
\ENDIF
\RETURN $(r_t, c_t, a_t)$
\end{algorithmic}
\end{algorithm}

\textbf{Fused generation mode (line~2).}
To avoid the repeated input cost, fused generation produces reasoning, conclusion, and action in a single pass. The total token consumption reduces to $\tau_{\text{in}}(x_t) + \tau_{\text{out}}(r_t, c_t, a_t)$, which eliminates the second input charge entirely. This provides additional savings on top of the cross-round compression from the conclusion chain, and avoids the need for a separate model, making it well suited for single-API deployments in edge and serverless~\cite{lou2024serverless} settings.

\textbf{Two-stage mode (lines~4--5).}
The main model reasons over $x_t$ and outputs $(r_t, a_t)$. A summary model (either the same model or a dedicated lightweight one) then extracts $c_t$ from $x_t \oplus r_t$. Let $\tau_{\text{in}}(\cdot)$ and $\tau_{\text{out}}(\cdot)$ denote the input and output token counts of an API call. The per-step cost of the two-stage mode involves two calls, so the total token consumption is roughly:
\begin{equation}
\tau_{\text{in}}(x_t) + \tau_{\text{out}}(r_t) + \tau_{\text{in}}(x_t \oplus r_t) + \tau_{\text{out}}(c_t).
\end{equation}
Because the second call re-sends $x_t$, the input cost is nearly doubled. This mode is suitable when a dedicated summary model is available or when separated prompts improve extraction quality.

The two modes are interchangeable. When a summary model is available, the two-stage mode can be used for better extraction quality. When only a single black-box API is accessible, fused generation keeps both cost and complexity low.

\section{Theoretical Analysis}
\label{sec:theory}
\subsection{Information-Theoretic Justification}

We characterize the difference between recursive summarization and the append-only rule from an information-theoretic perspective. Let $A$ denote the ground-truth answer as a random variable. The complete history up to round $t$ is $\mathcal{H}_t = [\mathcal{D}_1, r_1, \dots, \mathcal{D}_t, r_t]$. Since $\mathcal{S}_t$ is a deterministic function of $\mathcal{H}_t$, the chain $A \rightarrow \mathcal{H}_t \rightarrow \mathcal{S}_t$ forms a Markov chain, and by the data processing inequality:
\begin{equation}
I(\mathcal{S}_t;A\mid q)\le I(\mathcal{H}_t;A\mid q).
\end{equation}
The key question is not whether compression causes loss, but whether that loss accumulates across rounds.

\textbf{Recursive summarization.} Let $\sigma_t = f_{\text{rec}}(\sigma_{t-1}, \mathcal{D}_t)$ with $\sigma_0 = \emptyset$ denote a recursive summarization process, where $f_{\text{rec}}$ rewrites the previous summary together with new evidence into a single updated summary. Let $Z_k$ denote the fine-grained facts confirmed in round $k$. Assuming that later retrieval rounds do not re-retrieve the same facts as round $k$, the chain $Z_k \to \sigma_k \to \sigma_{k+1} \to \cdots \to \sigma_t$ forms a Markov chain. A cascade of the data processing inequality then yields:
\begin{equation}
I(\sigma_t;Z_k\mid q)\le I(\sigma_{t-1};Z_k\mid q)\le \cdots \le I(\sigma_k;Z_k\mid q).
\end{equation}
Because each rewriting step overwrites the previous text, the fidelity of early-round facts decays monotonically as the chain deepens.

\textbf{Append-only update.} ConCise follows the append-only rule $\mathcal{C}_t = \operatorname{Append}(\mathcal{C}_{t-1}, c_t)$. Because $c_k$ is retained verbatim in $\mathcal{C}_t$ for all $k \le t$, the conditional entropy $H(c_k \mid \mathcal{C}_t) = 0$. Therefore, for any $k < t$:
\begin{equation}
I(\mathcal{C}_t;c_k\mid q) = H(c_k\mid q).
\end{equation}
Round-$k$ conclusions do not decay in subsequent rounds. ConCise does not claim that single-step extraction is lossless; the core guarantee is that already-extracted conclusions are never recompressed, avoiding the cascading degradation of recursive summarization while maintaining intermediate-entity anchors under $O(N)$ payload growth.

\subsection{Complexity Analysis}

Under the cumulative input-token view, the full-context input at round $t$ is $x^{\text{full}}_t = q \oplus [\mathcal{D}_1, r_1, \dots, \mathcal{D}_{t-1}, r_{t-1}, \mathcal{D}_t]$. When the per-round token increment is bounded, the total input tokens across $N$ rounds grow as $O(N^2)$. ConCise replaces this with $x^{\text{concise}}_t = q \oplus \mathcal{C}_{t-1} \oplus \mathcal{D}_t$ (Eq.~(\ref{eq:concise_input})), where only the conclusion chain crosses round boundaries. Since each $c_t$ is much shorter than the pair $(\mathcal{D}_t, r_t)$ it replaces, cumulative growth reduces to $O(N)$.

Under the single-round prefill view with reusable KV caches, the incremental prefill per round is bounded by the per-round saving $\Delta_t$ (Eq.~(\ref{eq:delta})), showing a near-constant trend when $\Delta_t$ is bounded. Real cloud latency and memory are additionally affected by dynamic batching, prefix caching, and server-side scheduling.

\section{Experiment}
\label{sec:experiment}
\subsection{Baselines and Experimental Setup}

We select the following baselines that cover distinct approaches to multi-step RAG.
\begin{enumerate}
    \item \textbf{IRCoT}~\cite{trivedi2023ircot} is a representative prompt-driven multi-step RAG method. It interleaves retrieval with chain-of-thought reasoning and issues a new retrieval query at each step guided by the ongoing reasoning trace. Because IRCoT does not use reinforcement learning, it is a natural testbed for studying how state management affects reasoning quality.
    \item \textbf{Search-R1}~\cite{jin2025searchr1} is a representative agentic RAG method trained with reinforcement learning to decide when and what to search. It features tight search-reasoning coupling and serves as a more capable baseline. Including Search-R1 allows us to evaluate ConCise under a method with strong internal context handling.
\end{enumerate}
These two baselines represent the most widely adopted approaches in their respective categories, covering both prompt-driven and RL-trained paradigms. This pairing isolates how ConCise interacts with different levels of built-in state-management capability. Extending the comparison to additional methods is a valuable direction for future work.

The experiments cover twelve paired configurations across two datasets (HotpotQA~\cite{yang2018hotpotqa}, 2WikiMultihopQA~\cite{ho2020multihop}) and three models (GPT-5, Qwen2.5-72B, DeepSeek V3.2). All experiments run on UltraRAG~\cite{openbmb2025ultrarag} with the Tavily web API for retrieval. All models are accessed via remote API calls without local training (snapshot window: 2026-01-15 to 2026-02-28; no model version updates occur during the period). Retrieved documents are top-3 relevant snippets, and the maximum reasoning depth is set to four steps, matching the typical budget of mobile and edge service deployments. All experiments use the fused generation mode (Algorithm~\ref{alg:concise_update}, line~2). This paper focuses on two research questions (RQ):
\begin{enumerate}
\item RQ1: Can ConCise stably reduce context costs across different configurations?
\item RQ2: How does ConCise affect accuracy, and what factors determine the impact?
\end{enumerate}

\subsection{Metrics and Evaluation Criteria}

All comparisons are strictly paired under identical model, data, and retrieval settings. Baseline and ConCise configurations share the same UltraRAG implementation, Tavily retrieval interface, and model API call paths. ConCise modifies only the prompt template and historical concatenation strategy, minimizing implementation-induced bias.

We evaluate reasoning accuracy and token efficiency. Because output preferences differ across models and ConCise rewrites the prompt template, the output format may change without affecting correctness. String-overlap metrics would penalize such stylistic differences. We therefore employ LLM-as-Judge semantic-hit determination (SemAcc), where an independent LLM judge evaluates whether the predicted answer is semantically equivalent to the gold answer. The evaluation prompt instructs the judge to accept aliases, paraphrased answers, and correct partial answers while rejecting factually wrong responses. The same judge model and prompt are applied to both baseline and ConCise outputs under identical conditions to ensure fair comparison.

To evaluate efficiency, we track the average input token counts at steps 2, 3, and 4. The relative token saving rate at step $k$ is:
\begin{equation}
\begin{matrix}
s_k = \frac{t_k^{(b)} - t_k^{(c)}}{t_k^{(b)}} \times 100\%,
\end{matrix}
\end{equation}
where $t_k^{(b)}$ and $t_k^{(c)}$ denote the input token counts for the baseline and ConCise, respectively.

We supplement the efficiency evaluation with two service-side proxies, estimated network payload ($P_k$, in KB) and estimated input billing ($C_k$, in cents):
\begin{equation}
\begin{aligned}
P_k&=\frac{t_k\cdot b}{1024},\quad b=4\ \text{bytes/token},\\
C_k&=t_k\cdot \frac{p}{10^6}\cdot 100,
\end{aligned}
\end{equation}
where $p$ denotes the input token unit price (USD / 1M tokens). We adopt the GPT-5 standard input rate ($p=1.25$) from the official OpenAI API pricing page~\cite{openai2026pricing} (2026-02-28 snapshot) as a unified baseline. This conversion covers input tokens only, excluding output tokens and request JSON overhead.

Step~1 is single-step reasoning without ConCise intervention and is therefore excluded from token reduction metrics. Because we do not conduct key-value cache experiments, we do not report KV hit rates, prefill latency, or GPU memory usage. The maximum reasoning depth is four steps, so step~4 is synonymous with the final-round metric.

\subsection{Results and Analysis}
\label{sec:results}

Different baselines show varying sensitivities to state compression, so we adopt a stratified evaluation. Average results are reported by dataset and baseline method rather than merged across methods. Unless otherwise noted, Acc refers to SemAcc. TABLE~\ref{tab:agg_combined} summarizes the per-method averages (RQ1). The average step-4 token saving reaches 64.63\% (range: 49.6\%--74.1\%). Accuracy improves in five configuration groups and declines or remains flat in seven (one group shows $\Delta$Acc of 0.00, likely coincidental given single-run variability). Complete per-configuration details are provided in TABLE~\ref{tab:full_combined}. We summarize four findings from these results and analyze each below.

\begin{table*}[t]
\caption{Per-Method Average Results}
\label{tab:agg_combined}
\centering
\renewcommand{\arraystretch}{1.12}
\small
\setlength{\tabcolsep}{4pt}
\begin{tabular}{@{}ll cc c ccc cc@{}}
\toprule
& & \multicolumn{2}{c}{\textbf{Accuracy (\%)}} & & \multicolumn{3}{c}{\textbf{Token Saving (\%)}} & \multicolumn{2}{c}{\textbf{Service Proxy (Step4)}} \\
\cmidrule(lr){3-4} \cmidrule(l){6-8} \cmidrule(l){9-10}
\textbf{Dataset} & \textbf{Method} & Base & ConCise & $\Delta$\textbf{Acc} & Step 2 & Step 3 & Step 4 & \makecell{\textbf{Payload}\\\textbf{(KB)}} & \makecell{\textbf{Input Cost}\\\textbf{(cent)}} \\
\midrule
\multirow{2}{*}{\textit{2Wiki}}
 & Search-R1 & 72.39 & 71.10 & \down{1.29} & 23.90 & 48.22 & 59.10 & $10.87 \rightarrow 4.41$ & $0.348 \rightarrow 0.141$ \\
 & IRCoT      & 44.91 & 54.36 & \up{9.45}   & 45.77 & 63.15 & \textbf{69.16} & $9.48 \rightarrow 2.91$ & $0.303 \rightarrow 0.093$ \\
\midrule
\multirow{2}{*}{\textit{HotpotQA}}
 & Search-R1 & 61.95 & 61.09 & \down{0.86} & 39.56 & 59.87 & 63.45 & $12.56 \rightarrow 4.46$ & $0.402 \rightarrow 0.143$ \\
 & IRCoT      & 43.40 & 50.69 & \up{7.29}   & 39.27 & 62.73 & \textbf{66.81} & $9.23 \rightarrow 3.09$ & $0.295 \rightarrow 0.099$ \\
\bottomrule
\end{tabular}
\end{table*}

\begin{table*}[t]
\caption{Complete Detailed Results}
\label{tab:full_combined}
\centering
\renewcommand{\arraystretch}{1.12}
\small
\setlength{\tabcolsep}{4pt}
\begin{tabular}{@{}lll c c cc cc cc@{}}
\toprule
& & & & & \multicolumn{2}{c}{\textbf{Step 2}} & \multicolumn{2}{c}{\textbf{Step 3}} & \multicolumn{2}{c}{\textbf{Step 4}} \\
\cmidrule(lr){6-7} \cmidrule(lr){8-9} \cmidrule(l){10-11}
\textbf{Dataset} & \textbf{Method} & \textbf{Model} & \textbf{Acc (\%)} & $\boldsymbol{\Delta}$\textbf{Acc} & \textbf{Tok.} & \textbf{Save} & \textbf{Tok.} & \textbf{Save} & \textbf{Tok.} & \textbf{Save} \\
\midrule
\multirow{12}{*}{\rotatebox[origin=c]{90}{\textit{2WikiMultihopQA}}}
 & \multirow{6}{*}{\textit{Search-R1}}
 & GPT-5         & 82.83 & --    & 1228 & --    & 1890 & --    & 2589 & -- \\
 & & \cellcolor{conciserow}+ConCise  & \cellcolor{conciserow}82.40 & \cellcolor{conciserow}\down{0.43} & \cellcolor{conciserow}998  & \cellcolor{conciserow}18.7\% & \cellcolor{conciserow}1067 & \cellcolor{conciserow}43.5\% & \cellcolor{conciserow}1305 & \cellcolor{conciserow}\textbf{49.6\%} \\
\addlinespace[1.5pt]
 & & Qwen2.5-72B   & 59.23 & --    & 1363 & --    & 2072 & --    & 2835 & -- \\
 & & \cellcolor{conciserow}+ConCise  & \cellcolor{conciserow}59.23 & \cellcolor{conciserow}0.00  & \cellcolor{conciserow}1019 & \cellcolor{conciserow}25.3\% & \cellcolor{conciserow}1116 & \cellcolor{conciserow}46.2\% & \cellcolor{conciserow}1006 & \cellcolor{conciserow}\textbf{64.5\%} \\
\addlinespace[1.5pt]
 & & DeepSeek V3.2 & 75.11 & --    & 1437 & --    & 2157 & --    & 2921 & -- \\
 & & \cellcolor{conciserow}+ConCise  & \cellcolor{conciserow}71.67 & \cellcolor{conciserow}\down{3.43} & \cellcolor{conciserow}1039 & \cellcolor{conciserow}27.7\% & \cellcolor{conciserow}972  & \cellcolor{conciserow}55.0\% & \cellcolor{conciserow}1075 & \cellcolor{conciserow}\textbf{63.2\%} \\
\cmidrule(l){2-11}
 & \multirow{6}{*}{\textit{IRCoT}}
 & GPT-5         & 60.51 & --    & 1329 & --    & 1838 & --    & 2378 & -- \\
 & & \cellcolor{conciserow}+ConCise  & \cellcolor{conciserow}59.23 & \cellcolor{conciserow}\down{1.29} & \cellcolor{conciserow}723  & \cellcolor{conciserow}45.6\% & \cellcolor{conciserow}629  & \cellcolor{conciserow}65.8\% & \cellcolor{conciserow}683  & \cellcolor{conciserow}\textbf{71.3\%} \\
\addlinespace[1.5pt]
 & & Qwen2.5-72B   & 36.48 & --    & 1190 & --    & 1828 & --    & 2341 & -- \\
 & & \cellcolor{conciserow}+ConCise  & \cellcolor{conciserow}\textbf{52.79} & \cellcolor{conciserow}\up{16.31} & \cellcolor{conciserow}857  & \cellcolor{conciserow}28.0\% & \cellcolor{conciserow}769  & \cellcolor{conciserow}57.9\% & \cellcolor{conciserow}888  & \cellcolor{conciserow}\textbf{62.1\%} \\
\addlinespace[1.5pt]
 & & DeepSeek V3.2 & 37.74 & --    & 1333 & --    & 2000 & --    & 2563 & -- \\
 & & \cellcolor{conciserow}+ConCise  & \cellcolor{conciserow}51.07 & \cellcolor{conciserow}\up{13.34} & \cellcolor{conciserow}484  & \cellcolor{conciserow}63.7\% & \cellcolor{conciserow}685  & \cellcolor{conciserow}65.7\% & \cellcolor{conciserow}663  & \cellcolor{conciserow}\textbf{74.1\%} \\
\midrule
\multirow{12}{*}{\rotatebox[origin=c]{90}{\textit{HotpotQA}}}
 & \multirow{6}{*}{\textit{Search-R1}}
 & GPT-5         & 62.66 & --    & 1429 & --    & 3293 & --    & 3916 & -- \\
 & & \cellcolor{conciserow}+ConCise  & \cellcolor{conciserow}65.24 & \cellcolor{conciserow}\up{2.58}  & \cellcolor{conciserow}930  & \cellcolor{conciserow}34.9\% & \cellcolor{conciserow}1123 & \cellcolor{conciserow}65.9\% & \cellcolor{conciserow}1071 & \cellcolor{conciserow}\textbf{72.7\%} \\
\addlinespace[1.5pt]
 & & Qwen2.5-72B   & 58.80 & --    & 1394 & --    & 2043 & --    & 2600 & -- \\
 & & \cellcolor{conciserow}+ConCise  & \cellcolor{conciserow}57.51 & \cellcolor{conciserow}\down{1.29} & \cellcolor{conciserow}828  & \cellcolor{conciserow}40.6\% & \cellcolor{conciserow}957  & \cellcolor{conciserow}53.1\% & \cellcolor{conciserow}1087 & \cellcolor{conciserow}\textbf{58.2\%} \\
\addlinespace[1.5pt]
 & & DeepSeek V3.2 & 64.38 & --    & 1596 & --    & 2460 & --    & 3131 & -- \\
 & & \cellcolor{conciserow}+ConCise  & \cellcolor{conciserow}60.51 & \cellcolor{conciserow}\down{3.86} & \cellcolor{conciserow}907  & \cellcolor{conciserow}43.2\% & \cellcolor{conciserow}970  & \cellcolor{conciserow}60.6\% & \cellcolor{conciserow}1268 & \cellcolor{conciserow}\textbf{59.5\%} \\
\cmidrule(l){2-11}
 & \multirow{6}{*}{\textit{IRCoT}}
 & GPT-5         & 56.60 & --    & 1054 & --    & 1511 & --    & 2001 & -- \\
 & & \cellcolor{conciserow}+ConCise  & \cellcolor{conciserow}52.07 & \cellcolor{conciserow}\down{4.53} & \cellcolor{conciserow}681  & \cellcolor{conciserow}35.4\% & \cellcolor{conciserow}542  & \cellcolor{conciserow}64.1\% & \cellcolor{conciserow}655  & \cellcolor{conciserow}\textbf{67.3\%} \\
\addlinespace[1.5pt]
 & & Qwen2.5-72B   & 41.51 & --    & 1382 & --    & 2034 & --    & 2732 & -- \\
 & & \cellcolor{conciserow}+ConCise  & \cellcolor{conciserow}49.06 & \cellcolor{conciserow}\up{7.55}  & \cellcolor{conciserow}884  & \cellcolor{conciserow}36.0\% & \cellcolor{conciserow}740  & \cellcolor{conciserow}63.6\% & \cellcolor{conciserow}1052 & \cellcolor{conciserow}\textbf{61.5\%} \\
\addlinespace[1.5pt]
 & & DeepSeek V3.2 & 32.08 & --    & 1274 & --    & 1858 & --    & 2357 & -- \\
 & & \cellcolor{conciserow}+ConCise  & \cellcolor{conciserow}\textbf{50.94} & \cellcolor{conciserow}\up{18.86} & \cellcolor{conciserow}684  & \cellcolor{conciserow}\textbf{46.3\%} & \cellcolor{conciserow}735  & \cellcolor{conciserow}60.5\% & \cellcolor{conciserow}668  & \cellcolor{conciserow}\textbf{71.7\%} \\
\bottomrule
\end{tabular}
\end{table*}

\textbf{Finding 1 (RQ1): Token savings are substantial and consistent.} All four dataset--method groups show stable savings at steps~3 and~4, with a clear depth amplification effect. Within the four-step setting, the saving rate generally increases as reasoning rounds deepen (step~2 $<$ step~3 $<$ step~4 in 10 of 12 configurations; two configurations show a slight decrease from step~3 to step~4). This matches the trend in Fig.~\ref{fig:token_trend}. As rounds increase, the full-context approach accumulates more historical burden while ConCise appends only conclusion items, so the relative cost gap widens with each round.

\textbf{Finding 2 (RQ2): Quality impact depends on the baseline method.} Accuracy trends diverge between the two baselines. On 2Wiki, IRCoT averages $+9.45$ pp while Search-R1 averages $-1.29$ pp. On HotpotQA, IRCoT averages $+7.29$ pp while Search-R1 averages $-0.86$ pp. Stratified reporting by method captures this divergence more faithfully than a single merged average.

We attribute this divergence to baseline sensitivity to long-context noise. IRCoT is a prompt-driven method that is more susceptible to redundancy and evidence drift. The conclusion chain provides a compact factual summary at each step, focusing subsequent reasoning on confirmed sub-conclusions and current evidence. Case~1 in Section~\ref{sec:cases} provides direct evidence: the conclusion chain preserves entity relationships from step~1 through all subsequent steps, preventing the attention drift that causes the baseline to converge on a wrong answer. Search-R1, by contrast, already has tight search-reasoning coupling through reinforcement learning. In such configurations, structured compression may discard soft clues such as weakly associated entities or fine-grained constraints. Search-R1 degradation cases concentrate on questions requiring fine-grained temporal or comparative reasoning (e.g., Case~2), where the conclusion loses numeric details that the RL policy has learned to use. Verifying this hypothesis with intermediate-step attention distributions is left to future work.

This divergence does not undermine the value of ConCise. For strong baselines, the primary benefit is resource efficiency; for methods more susceptible to long-context interference, quality improvements may emerge as an additional benefit.

\textbf{Finding 3 (RQ2): Gains are driven by state management gaps rather than model scale.} Aggregated by method across three models and two datasets, the Search-R1 groups average $\Delta$Acc of $-1.07$ pp while the IRCoT groups average $+8.37$ pp. For Qwen2.5-72B, IRCoT results on both datasets are positive ($+16.31$ pp / $+7.55$ pp), while Search-R1 results are approximately flat. DeepSeek V3.2 shows the same pattern: IRCoT is positive ($+13.34$ pp / $+18.86$ pp) while Search-R1 is negative on both datasets. GPT-5 varies across methods, showing task dependence. These observations indicate that the accuracy impact depends on how well the method already manages long-context state, not on model scale alone. When a strong model such as GPT-5 already extracts key information from long contexts, ConCise reduces overhead but may also remove marginally useful evidence, causing minor accuracy drops.

\textbf{Finding 4: Three failure modes are identified.} First, missing information: incomplete conclusion extraction loses key details needed for subsequent reasoning. Second, fine-grained loss: numbers, timestamps, and lists degrade more easily during compression because they carry high information density in few tokens. Third, error propagation: an incorrect conclusion written in an early round is carried verbatim into all subsequent rounds and may never be corrected. These observations indicate that the upper bound of ConCise depends on conclusion quality. We analyze concrete instances of these failure modes in the case studies (Section~\ref{sec:cases}).

\subsection{Deployment Implications}
\label{sec:deployment}

The depth amplification effect (Finding~1) suggests that ConCise offers higher savings in longer reasoning chains. This is relevant for edge agents and weak-network scenarios~\cite{wang2025edgesurvey} where multi-step evidence verification faces latency and memory pressure. Because ConCise works at the prompt-concatenation level without additional model services, it keeps a single remote LLM API call chain and preserves the thin-client property of Web and edge deployments. ConCise and system-level KV cache optimizations operate at different layers (semantic state versus inference engine) and can in principle be combined, though this paper does not provide joint empirical evidence.

The token counts reported in the experiments are recorded from API responses and are exact values, not approximations. The payload and billing proxies in TABLE~\ref{tab:agg_combined} are linear conversions of these counts (Eqs.~4--5), so the proportional savings translate directly to actual cost reduction under the same pricing model.

\section{Case Studies}
\label{sec:cases}

\textbf{Case 1: Intermediate-state hijacking (IRCoT).}
The question asks which prime minister born in Stanmore served during the tenure of John Gorton. The golden answer is Harold Holt. The baseline outputs Robert Menzies (incorrect), while ConCise outputs Harold Holt (correct). The baseline does not lack evidence; rather, an intermediate candidate entity is amplified repeatedly in subsequent rounds and hijacks the final answer. ConCise retains the relationship constraints through the conclusion chain and converges correctly. Table~\ref{tab:case_a_raw} shows the output fragment comparison. This case illustrates the attention-anchor benefit described in Finding~2. By replacing accumulated raw text with structured conclusions, ConCise prevents repeated mentions of a wrong candidate from dominating the model's attention.

\begin{table*}[t]
\caption{Case 1 (IRCoT): Raw Output Snippet Comparison}
\label{tab:case_a_raw}
\centering
\renewcommand{\arraystretch}{1.15}
\small
\begin{tabular}{@{}c p{6.8cm} p{6.8cm}@{}}
\toprule
\textbf{Round} & \textbf{Baseline} & \textbf{+ConCise} \\
\midrule
Step 1 & \dots Among these, Robert Menzies was born in Stanmore. So the answer is: Robert Menzies. & C1: John Gorton served as a minister under Harold Holt, who was born in Stanmore. \\
\addlinespace[2pt]
Step 2 & Continues to output Robert Menzies & \dots we can conclude \dots Harold Holt \dots \\
\addlinespace[2pt]
Step 3 & Still reasons around Robert Menzies & Relationship anchors remain stable despite local irrelevant information \\
\addlinespace[2pt]
Step 4 & \textbf{Final:} Robert Menzies \textcolor{red!70!black}{(\ding{55}\,wrong)} & \textbf{Final:} Harold Holt \textcolor{green!60!black}{(\ding{51}\,correct)} \\
\bottomrule
\end{tabular}

\vspace{0.6em}

\caption{Case 2 (Search-R1): Conclusion-Chain Evolution and Risk}
\label{tab:case_b_raw}
\centering
\renewcommand{\arraystretch}{1.15}
\small
{\setlength{\tabcolsep}{2.8pt}
\begin{tabular}{@{}c p{8.6cm} p{5.2cm}@{}}
\toprule
\textbf{Round} & \textbf{ConCise Conclusion Fragment} & \textbf{Risk Annotation} \\
\midrule
C1 & \dots cannot determine which died first \dots & \textcolor{red!70!black}{Writes uncertain state into the chain} \\
\addlinespace[2pt]
C2 & \dots cannot determine which died first \dots & \textcolor{red!70!black}{State not overturned by new evidence} \\
\addlinespace[2pt]
C3 & \dots cannot determine who died first \dots & \textcolor{red!70!black}{Uncertainty structurally solidified} \\
\bottomrule
\end{tabular}
}
\end{table*}

\textbf{Case 2: Irreversible compression trap (Search-R1).}
The question asks which film director died earlier between K\"{o}pekler Adas{\i} and Twisted Mistress. The golden answer is Twisted Mistress. The baseline answers correctly, while ConCise answers incorrectly. The key issue is that a low-confidence indeterminate state is written into the chain and solidified in early rounds. Subsequent evidence fails to overturn it, leading to incorrect convergence. Table~\ref{tab:case_b_raw} illustrates the chain evolution. This case demonstrates the error propagation failure mode described in Finding~4. Once an uncertain conclusion is committed to the append-only chain, it is carried verbatim into all future rounds and prevents self-correction.

\textbf{Case 3: Initial overhead versus asymptotic savings (Search-R1).}
The question asks where the director of Nalini By Day, Nancy By Night graduated. The golden answer is Mount Holyoke College. The baseline fails to converge (remaining in a search action), while ConCise converges correctly. Approximate main prompt tokens (estimated via character count / 4, as exact tokenizer counts are not recorded for this sample) are $[922, 2000, 3076, 4151]$ for the baseline and $[1458, 1458, 1458, 1458]$ for ConCise across steps~1--4. The cumulative totals are 10{,}149 versus 5{,}832, a reduction of about 42.5\%. This illustrates the tradeoff between a higher step-1 constant (conclusion-chain overhead) and a lower growth slope from step~2 onward, consistent with the $O(N)$ versus $O(N^2)$ analysis in Section~\ref{sec:theory}.

\textbf{Mitigation directions.} The three case studies illustrate both the attention-anchor benefit (Case~1) and the irreversible compression risk (Case~2) of the append-only design. The append-only rule trades correctability for fidelity: once a conclusion is committed, it is preserved exactly but cannot be revised. This is a fundamental tradeoff, and the error propagation risk (Finding~4) is its main cost. To address this, three directions merit investigation. (i)~Delayed chain insertion, which postpones low-confidence conclusions to avoid hardening uncertain states. For example, a conclusion containing hedging phrases such as ``cannot determine'' could be held in a buffer rather than committed. (ii)~Evidence window, which retains raw evidence from the most recent $k$ rounds (e.g., $k=2$) for refutation and backtracking. (iii)~Degradation trigger, which activates reduced-compression mode when the conclusion chain fails to converge for two consecutive rounds. We leave these extensions to future work.

\section{Conclusion}
\label{sec:conclusion}

This paper presented ConCise, a training-free state management method that replaces raw-text accumulation with an append-only conclusion chain, compressing cumulative context growth in multi-step RAG from $O(N^2)$ to approximately $O(N)$. A fused generation mode further removed the repeated input cost of the two-call workflow. Across twelve paired configurations, ConCise achieved 64.63\% average token savings at step~4 while keeping accuracy loss within acceptable bounds. As a prompt-level component requiring no model modification, ConCise is readily deployable in API-native and edge Web service settings. The current evaluation is limited to four reasoning steps with two baseline methods and does not compare fused and two-stage generation modes. Multi-run evaluation with variance reporting would further strengthen the results. Future work includes mitigating the hardening of early uncertain conclusions through delayed insertion or a sliding evidence window, evaluating deeper reasoning chains, comparing generation modes, extending the baseline comparison to additional state-management methods, and joint evaluation with system-level KV cache optimizations.

\bibliographystyle{IEEEtran}
\bibliography{references}

\end{document}